\documentclass[a4paper]{article}

\pdfoutput=1

\usepackage{INTERSPEECH2018}

\usepackage{bm}
\usepackage{tikz}
\usepackage{pgfplots}

\usepackage{hyperref}
\usepackage[T1]{fontenc}
\usepackage[utf8]{inputenc}
\usepackage{pgfplots}
\usepackage{grffile}
\pgfplotsset{compat=newest}
\usetikzlibrary{plotmarks}
\usetikzlibrary{arrows.meta}
\usepgfplotslibrary{patchplots}

\usetikzlibrary{arrows}

\title{Speaker-independent raw waveform model for glottal excitation}
\name{Lauri Juvela$^1$, Vassilis Tsiaras$^2$,  Bajibabu Bollepalli$^1$, \\ Manu Airaksinen$^1$, Junichi Yamagishi$^3$, Paavo Alku$^1$}
\address{
  $^1$Aalto University, Finland\\
  $^2$University of Crete, Greece\\
  $^3$National Institute of Informatics, Japan}
\email{lauri.juvela@aalto.fi, tsiaras@csd.uoc.gr}

\ninept

\begin{document}

\maketitle
\begin{abstract}
%Recently, there has been an increasing speech technology research interest in 
%Recently, speech technology research has shown a growing interest in 
%Recent speech technology research has focused on
Recent speech technology research has seen a growing interest in
using WaveNets as statistical vocoders, i.e., generating speech waveforms from acoustic features. These models have been shown to improve the generated speech quality over classical vocoders in many tasks, such as text-to-speech synthesis and voice conversion. Furthermore, conditioning WaveNets with acoustic features allows sharing the waveform generator model across multiple speakers without additional speaker codes. 
However, multi-speaker WaveNet models require large amounts of training data and computation to cover the entire acoustic space. This paper proposes leveraging  the source-filter model of speech production to more effectively train a speaker-independent waveform generator with limited resources. 
We present a multi-speaker 'GlotNet' vocoder, which utilizes a WaveNet to generate glottal excitation waveforms, which are then used to excite the corresponding vocal tract filter to produce speech. Listening tests show that the proposed model performs favourably to a direct WaveNet vocoder trained with the same model architecture and data. 

% 157/200 words
 
\end{abstract}

\section{Introduction}

Recently, there has been a growing interest in WaveNet-based waveform generation in speech applications due to the high quality of generated speech. 
While the first WaveNet text-to-speech (TTS) model used linguistic features and fundamental frequency (F0) from an existing statistical parametric speech synthesis (SPSS) system \cite{oord2016-wavenet}, there seems to be a shift in focus towards using WaveNets as statistical vocoders. In the statistical vocoder approach, a WaveNet is conditioned with some acoustic features, such as mel filterbank energies \cite{Arik2017-deepvoice, Shen2018-tacotron2}, or mel-generalized cepstrum (MGC) coefficients and F0 \cite{Tamamori2017-wavenet-vocoder}. 
In context of TTS, high-quality systems have been built by separately training a WaveNet vocoder and  a text-to-acoustic-features model, where the latter can be an end-to-end attention-based neural net \cite{Arik2017-deepvoice, Shen2018-tacotron2} or a more conventional frame-aligned SPSS system \cite{Wang2018-comparison-of-waveform-generation}.

A clear benefit of acoustically conditioned WaveNets is that the same waveform generator model can be shared between multiple speakers, provided that the acoustic features contain sufficient information to capture the speaker identity.  
For example, multi-speaker WaveNets have been successfully conditioned on  low-bitrate speech codec parameters  \cite{Kleijn2018-wavenet-low-rate-coding}, as well as on acoustic parameters typically used in parametric TTS (MGC, F0) \cite{Hayashi2017-multispeaker-wavenet-vocoder}.
Furthermore, previous research found no added benefit from using speaker codes to supplement the acoustic features \cite{Hayashi2017-multispeaker-wavenet-vocoder}, which suggests that the acoustic features themselves can be sufficient for high-quality speaker-independent waveform generation. However, training large-scale speaker-independent models that cover the acoustic space for various unseen speakers is expected to be costly in terms of data and computation. This problem can be mitigated by leveraging knowledge of the human speech production mechanism to reduce the data variability in speech.

Before WaveNets, waveform synthesis with neural networks has been applied, using simple fully connected networks \cite{juvela2016a-high-pitched-excitation, airaksinen2016glottdnn}, to glottal excitations, i.e., time-domain signals corresponding to the volume velocity waveform generated by the vocal folds in the human speech production mechanism. In this approach, the target waveform is a glottal excitation signal  estimated from speech using glottal inverse filtering (GIF), specifically quasi-closed phase (QCP) analysis \cite{Airaksinen2014}. GIF decomposes a speech signal into a vocal tract filter and a glottal source, effectively removing the vocal tract resonances from speech \cite{Alku2011}. %[REF].
Due to the absence of vocal tract resonances, the glottal excitation signal is more elementary than the speech pressure signal, and thus easier to model and synthesize with simple neural nets. 
%Before WaveNets, waveform synthesis with neural networks has been applied to glottal excitation signals, using simple fully connected networks %\cite{juvela2016a-high-pitched-excitation, airaksinen2016glottdnn}. 
%In this approach, the target waveform is a glottal excitation signal  estimated from speech using glottal inverse filtering (GIF) techniques, specifically %quasi-closed phase (QCP) analysis \cite{Airaksinen2014}.  
%Glottal inverse filtering decomposes a speech signal into a vocal tract filter and a glottal source excitation, effectively removing the vocal tract resonances %from speech. The residual glottal excitation signal is more elementary than the speech pressure signal, and has thus been easier to synthesize with simple neural %nets. 
%
Similarly to the emerging WaveNet vocoders, previous glottal waveform synthesis models have mostly used acoustic features as the conditioning input. However, in contrast to the sample-by-sample generation of WaveNets, these glottal waveform models used a pitch synchronous frame-based waveform representation. While this representation facilitates learning (and is applicable to parallel inference), the approach is sensitive to pitch-tracking errors and is limited to producing voiced speech only. Furthermore, these models were trained using least-squares regression, which does not allow true stochastic sampling from the learned distribution.
More recently, generative adversarial networks have been applied to the task to enable stochastic generation  \cite{Bollepalli2017-gan-glottal-excitation, juvela2018-synthesis-from-mfcc}, but these models are still constrained by the pitch synchronous windowing scheme. 

With WaveNets now available, it is natural to extend the generation of glottal excitation signals to utilize WaveNet-like models. This paper presents GlotNet, a speaker-independent neural waveform generator explicitly based on the source-filter model of speech production: a WaveNet conditioned on acoustic features generates a glottal source signal, which is then used to excite an all-pole vocal tract filter. The proposed system is compared with a direct speech pressure signal WaveNet vocoder trained using the same model architecture, acoustic conditioning and dataset. Additionally, we propose a simple but effective method for including a non-causal look-ahead into the acoustic conditioning. 
Although the paper scope is limited to copy-synthesis (i.e., natural acoustic features are used at test time), the proposed method should interface well with the ever-improving acoustic models in TTS systems.
%
%Listening test results show that the GlotNet achieves significantly better speaker similarity ratings on both seen and unseen speakers, and is clearly preferred in pairwise CCR quality comparison test.

The paper is structured as follows: Section \ref{sec:wavegen_models} describes the waveform generator models, while the experiments and evaluation are described in Section \ref{sec:experiments}. We discuss the results in Section \ref{sec:discussion} and conclude in Section \ref{sec:conclusions}.

%\cite{Juvela2018-glotnet-tts}

\section{Waveform generator models}
\label{sec:wavegen_models}

\begin{figure}[thb]
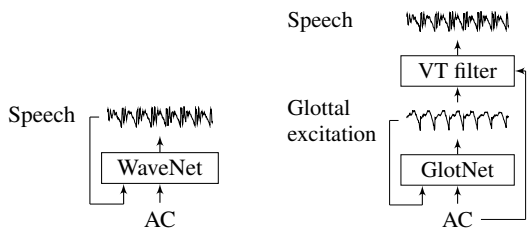

    %\centering
    \begin{minipage}[b][][t]{.45\linewidth}
    \vfill
     %\centering
    {
    %Large
    \resizebox{!}{0.5\linewidth}{\begin{tikzpicture}
[align=center,node distance=2.2em]
\tikzstyle{rectnode}=[ rectangle, minimum width=5em, minimum height=1.0em, draw]

\tikzstyle{line}= [-latex', line width=0.15mm]

% main nodes

\node[] (ac) {AC};
\node[rectnode, above of=ac] (wave_model) {WaveNet};
\node[ above of = wave_model, inner sep=3pt] (wave_output) {\resizebox{4.5em}{!}{% This file was created by matlab2tikz.
\begin{tikzpicture}

\begin{axis}[%
width=9.322in,
height=1.864in,
at={(1.564in,0.613in)},
scale only axis,
xmin=0,
xmax=1,
ymin=-0.54269,
ymax=0.45731,
axis line style={draw=none},
ticks=none
]
\addplot [color=black, line width=3.0pt, forget plot]
  table[row sep=crcr]{%
0	0.19397\\
0.002681	0.26002\\
0.0053619	0.34282\\
0.0080429	0.27869\\
0.010724	-0.061685\\
0.013405	-0.18599\\
0.016086	-0.20552\\
0.018767	-0.10138\\
0.021448	-0.059124\\
0.024129	0.083428\\
0.02681	0.24444\\
0.029491	0.28243\\
0.032172	0.12269\\
0.034853	0.11373\\
0.037534	0.047684\\
0.040214	-0.06702\\
0.042895	-0.089107\\
0.045576	0.017381\\
0.048257	0.043095\\
0.050938	0.08727\\
0.053619	0.14606\\
0.0563	0.27837\\
0.058981	0.28563\\
0.061662	0.28168\\
0.064343	0.28296\\
0.067024	0.20923\\
0.069705	0.089724\\
0.072386	0.051311\\
0.075067	0.075959\\
0.077748	0.13219\\
0.080429	0.1562\\
0.08311	0.18021\\
0.085791	0.16089\\
0.088472	0.054512\\
0.091153	-0.032662\\
0.093834	-0.018044\\
0.096515	-0.061365\\
0.099196	-0.12048\\
0.10188	-0.088787\\
0.10456	-0.078864\\
0.10724	-0.16316\\
0.10992	-0.17575\\
0.1126	-0.16497\\
0.11528	-0.15185\\
0.11796	-0.24958\\
0.12064	-0.27167\\
0.12332	-0.25353\\
0.12601	-0.34263\\
0.12869	-0.48635\\
0.13137	-0.51122\\
0.13405	-0.54269\\
0.13673	-0.14811\\
0.13941	0.1896\\
0.14209	0.38198\\
0.14477	0.26898\\
0.14745	0.40182\\
0.15013	0.1132\\
0.15282	-0.26495\\
0.1555	-0.31329\\
0.15818	-0.11642\\
0.16086	-0.15355\\
0.16354	0.015567\\
0.16622	0.22651\\
0.1689	0.33375\\
0.17158	0.19408\\
0.17426	0.10744\\
0.17694	0.046723\\
0.17962	-0.037997\\
0.18231	-0.17415\\
0.18499	-0.082812\\
0.18767	0.031572\\
0.19035	0.081081\\
0.19303	0.097086\\
0.19571	0.21744\\
0.19839	0.23292\\
0.20107	0.18362\\
0.20375	0.16708\\
0.20643	0.22097\\
0.20912	0.18181\\
0.2118	0.11192\\
0.21448	0.12654\\
0.21716	0.17786\\
0.21984	0.15684\\
0.22252	0.1515\\
0.2252	0.18565\\
0.22788	0.15684\\
0.23056	0.061875\\
0.23324	0.042775\\
0.23592	-0.0099349\\
0.23861	-0.066593\\
0.24129	-0.078543\\
0.24397	-0.054429\\
0.24665	-0.084946\\
0.24933	-0.12507\\
0.25201	-0.1463\\
0.25469	-0.16422\\
0.25737	-0.2227\\
0.26005	-0.29227\\
0.26273	-0.33228\\
0.26542	-0.34188\\
0.2681	-0.36429\\
0.27078	-0.4297\\
0.27346	-0.50759\\
0.27614	-0.52242\\
0.27882	0.18511\\
0.2815	0.22651\\
0.28418	0.27848\\
0.28686	0.28456\\
0.28954	0.33343\\
0.29223	-0.075663\\
0.29491	-0.35095\\
0.29759	-0.32086\\
0.30027	-0.043546\\
0.30295	-0.13787\\
0.30563	0.096126\\
0.30831	0.3361\\
0.31099	0.37568\\
0.31367	0.10616\\
0.31635	0.081401\\
0.31903	-0.013563\\
0.32172	-0.11162\\
0.3244	-0.23827\\
0.32708	-0.049948\\
0.32976	0.072225\\
0.33244	0.12248\\
0.33512	0.1148\\
0.3378	0.24732\\
0.34048	0.18383\\
0.34316	0.094845\\
0.34584	0.049071\\
0.34853	0.13454\\
0.35121	0.15268\\
0.35389	0.18586\\
0.35657	0.21051\\
0.35925	0.26343\\
0.36193	0.18778\\
0.36461	0.1403\\
0.36729	0.13176\\
0.36997	0.11896\\
0.37265	0.07852\\
0.37534	0.10178\\
0.37802	0.043095\\
0.3807	-0.030742\\
0.38338	-0.080251\\
0.38606	-0.065099\\
0.38874	-0.11845\\
0.39142	-0.13894\\
0.3941	-0.13381\\
0.39678	-0.12677\\
0.39946	-0.22291\\
0.40214	-0.26655\\
0.40483	-0.31158\\
0.40751	-0.30987\\
0.41019	-0.39022\\
0.41287	-0.45637\\
0.41555	-0.52221\\
0.41823	-0.43162\\
0.42091	0.32831\\
0.42359	0.27944\\
0.42627	0.22907\\
0.42895	0.41228\\
0.43164	0.26023\\
0.43432	-0.22536\\
0.437	-0.39523\\
0.43968	-0.35063\\
0.44236	-0.086973\\
0.44504	-0.13221\\
0.44772	0.18138\\
0.4504	0.44098\\
0.45308	0.39926\\
0.45576	0.087056\\
0.45845	0.062302\\
0.46113	-0.11397\\
0.46381	-0.21704\\
0.46649	-0.30603\\
0.46917	-0.054536\\
0.47185	0.10658\\
0.47453	0.15791\\
0.47721	0.16505\\
0.47989	0.28744\\
0.48257	0.15225\\
0.48525	0.053659\\
0.48794	0.048217\\
0.49062	0.12611\\
0.4933	0.11128\\
0.49598	0.12088\\
0.49866	0.19269\\
0.50134	0.2597\\
0.50402	0.15737\\
0.5067	0.13123\\
0.50938	0.16143\\
0.51206	0.093458\\
0.51475	0.060701\\
0.51743	0.062088\\
0.52011	0.017274\\
0.52279	-0.074062\\
0.52547	-0.084199\\
0.52815	-0.058377\\
0.53083	-0.073955\\
0.53351	-0.15974\\
0.53619	-0.15569\\
0.53887	-0.15046\\
0.54155	-0.24692\\
0.54424	-0.35287\\
0.54692	-0.33719\\
0.5496	-0.32406\\
0.55228	-0.40878\\
0.55496	-0.47056\\
0.55764	-0.4393\\
0.56032	0.36843\\
0.563	0.23793\\
0.56568	0.30184\\
0.56836	0.2773\\
0.57105	0.3043\\
0.57373	-0.10106\\
0.57641	-0.46469\\
0.57909	-0.36162\\
0.58177	-0.000972\\
0.58445	-0.12891\\
0.58713	0.17199\\
0.58981	0.40374\\
0.59249	0.42978\\
0.59517	0.16452\\
0.59786	-0.035436\\
0.60054	-0.090387\\
0.60322	-0.14918\\
0.6059	-0.31457\\
0.60858	-0.097003\\
0.61126	0.15566\\
0.61394	0.17188\\
0.61662	0.18864\\
0.6193	0.2088\\
0.62198	0.19109\\
0.62466	0.054619\\
0.62735	-0.019004\\
0.63003	0.078627\\
0.63271	0.14926\\
0.63539	0.087696\\
0.63807	0.18074\\
0.64075	0.22662\\
0.64343	0.19621\\
0.64611	0.097299\\
0.64879	0.11768\\
0.65147	0.093031\\
0.65416	0.067957\\
0.65684	0.02581\\
0.65952	0.02741\\
0.6622	-0.048561\\
0.66488	-0.092094\\
0.66756	-0.1034\\
0.67024	-0.12549\\
0.67292	-0.16785\\
0.6756	-0.19751\\
0.67828	-0.1766\\
0.68097	-0.2482\\
0.68365	-0.33025\\
0.68633	-0.35426\\
0.68901	-0.31008\\
0.69169	-0.42521\\
0.69437	-0.4887\\
0.69705	-0.27445\\
0.69973	0.41879\\
0.70241	0.29267\\
0.70509	0.13443\\
0.70777	0.39393\\
0.71046	0.21147\\
0.71314	-0.26591\\
0.71582	-0.40249\\
0.7185	-0.33004\\
0.72118	0.043202\\
0.72386	-0.073635\\
0.72654	0.17615\\
0.72922	0.45731\\
0.7319	0.39009\\
0.73458	0.071371\\
0.73727	-0.029888\\
0.73995	-0.12005\\
0.74263	-0.15836\\
0.74531	-0.28298\\
0.74799	-0.063605\\
0.75067	0.20613\\
0.75335	0.18896\\
0.75603	0.1912\\
0.75871	0.22854\\
0.76139	0.1833\\
0.76408	0.069771\\
0.76676	-0.0042797\\
0.76944	0.084922\\
0.77212	0.14777\\
0.7748	0.095486\\
0.77748	0.17455\\
0.78016	0.2263\\
0.78284	0.19418\\
0.78552	0.10861\\
0.7882	0.098046\\
0.79088	0.090791\\
0.79357	0.086523\\
0.79625	0.0044697\\
0.79893	0.0041496\\
0.80161	-0.051015\\
0.80429	-0.072782\\
0.80697	-0.097963\\
0.80965	-0.14459\\
0.81233	-0.14512\\
0.81501	-0.17916\\
0.81769	-0.20072\\
0.82038	-0.27178\\
0.82306	-0.31435\\
0.82574	-0.36674\\
0.82842	-0.331\\
0.8311	-0.39534\\
0.83378	-0.44773\\
0.83646	-0.18855\\
0.83914	0.37611\\
0.84182	0.30355\\
0.8445	0.13166\\
0.84718	0.33354\\
0.84987	0.1785\\
0.85255	-0.25428\\
0.85523	-0.38574\\
0.85791	-0.31414\\
0.86059	0.039361\\
0.86327	-0.051335\\
0.86595	0.14884\\
0.86863	0.40321\\
0.87131	0.3552\\
0.87399	0.12291\\
0.87668	-0.054216\\
0.87936	-0.10671\\
0.88204	-0.12784\\
0.88472	-0.23966\\
0.8874	-0.11162\\
0.89008	0.14425\\
0.89276	0.18768\\
0.89544	0.20251\\
0.89812	0.17423\\
0.9008	0.22726\\
0.90349	0.15524\\
0.90617	0.013753\\
0.90885	0.066143\\
0.91153	0.11266\\
0.91421	0.099007\\
0.91689	0.11117\\
0.91957	0.17135\\
0.92225	0.20507\\
0.92493	0.15652\\
0.92761	0.075106\\
0.93029	0.08919\\
0.93298	0.075746\\
0.93566	0.0063904\\
0.93834	-0.052508\\
0.94102	-0.056883\\
0.9437	-0.03597\\
0.94638	-0.085266\\
0.94906	-0.14182\\
0.95174	-0.13627\\
0.95442	-0.15409\\
0.9571	-0.17735\\
0.95979	-0.23198\\
0.96247	-0.26324\\
0.96515	-0.28256\\
0.96783	-0.30998\\
0.97051	-0.36557\\
0.97319	-0.37037\\
0.97587	-0.43642\\
0.97855	0.095379\\
0.98123	0.21179\\
0.98391	0.31881\\
0.9866	0.20197\\
0.98928	0.25586\\
0.99196	0.12387\\
0.99464	-0.26804\\
0.99732	-0.3071\\
1	-0.22814\\
};
\end{axis}
\end{tikzpicture}%
}};

%main connects
\draw[line] (ac) -- (wave_model);
\draw[line] (wave_model) -- (wave_output);

\node[left of=wave_output, node distance=4.0em, text width=5em, align=flush left] (speech) {Speech};

% loop for signal
\coordinate[ left of=wave_output, node distance = 3em] (left_of_wave) ;
\coordinate[ below left of=wave_model] (wave_feedback);
\coordinate (a) at ( left_of_wave |- wave_feedback);

\draw[] (wave_output) -- (left_of_wave);
\draw[] (left_of_wave) -- (a);
\draw[] (a) -- (wave_feedback);
\draw[line] (wave_feedback) -- (wave_feedback |- wave_model.south);

\end{tikzpicture}}
    }
    \end{minipage}%
    \hspace{0pt}
    \begin{minipage}[b][][t]{.45\linewidth}
    \vfill
     %\centering
    {
    %\Large
    \resizebox{!}{0.85\linewidth}{\begin{tikzpicture}
[align=center,node distance=2.2em]
\tikzstyle{rectnode}=[ rectangle, minimum width=5em, minimum height=1.0em, draw]
\tikzstyle{line}= [-latex', line width=0.15mm]

% main nodes

\node[] (ac) {AC};
\node[rectnode, above of=ac] (wave_model) {GlotNet};
\node[ above of = wave_model, inner sep=3pt] (wave_output) {\resizebox{4.5em}{!}{\input{figures/excitation.tikz}}};
\node[rectnode, above of=wave_output] (filter) {VT filter};
\node[ above of = filter, inner sep=3pt] (speech_output) {\resizebox{4.5em}{!}{\input{figures/signal.tikz}}};

% labels
\node[left of=wave_output, node distance=5.0em, text width=5em, align=flush left] (glot) {Glottal excitation};
\node[left of=speech_output, node distance=5.0em, text width=5em, align=flush left] (speech) {Speech};

%main connects
\draw[line] (ac) -- (wave_model);
\draw[line] (wave_model) -- (wave_output);
\draw[line] (wave_output) -- (filter);
\draw[line] (filter) -- (speech_output);

% loop to filter
\coordinate[ right of=ac, node distance = 3em] (right_of_ac) ;
\coordinate (right_of_filter) at ( filter -| right_of_ac);
\draw[] (ac) -- (right_of_ac);
\draw[] (right_of_ac) -- (right_of_filter);
\draw[line] (right_of_filter) -- (filter);

% loop for signal
\coordinate[ left of=wave_output, node distance = 3em] (left_of_wave) ;
\coordinate[ below left of=wave_model] (wave_feedback);
\coordinate (a) at ( left_of_wave |- wave_feedback);

\draw[] (wave_output) -- (left_of_wave);
\draw[] (left_of_wave) -- (a);
\draw[] (a) -- (wave_feedback);
\draw[line] (wave_feedback) -- (wave_feedback |- wave_model.south);

\end{tikzpicture}}
    }
    \end{minipage}
    %\vspace{-10pt}
    \caption{ WaveNet vocoder (left) uses acoustic features (AC) and past signal samples to generate the next speech sample. In contrast, GlotNet (right) operates on the more simplistic glottal excitation signal, which is filtered by a vocal tract (VT) filter already parametrized in the acoustic features. }
    \label{fig:simple_blockdiagram}
\end{figure}

An overview of the WaveNet and GlotNet vocoders is shown in Fig.~\ref{fig:simple_blockdiagram}.  While a WaveNet vocoder learns a non-linear autoregressive (AR) model to predict next signal sample from previous samples signal and time-varying acoustic features, a GlotNet operates on a more simplistic glottal excitation signal. The excitation signal is then passed through an all-pole vocal tract (VT) filter to produce speech waveforms. % This source-filter decomposition is motivated by 
The GlotNet model for the speech signal $x_n$ can be viewed as a mixture of a low-order linear AR process (VT filter) and a non-linear residual excitation process $e_n$ (glottal source)
% \begin{equation}
%     x_t = \sum_{k=1}^P a_k x_{t-k} + e_t(e_{(t-R):(t-1)}, h_n) ,
% \end{equation}
\begin{equation}
    x_n = \sum_{k=1}^P a_k x_{n-k} + e_n ,
\end{equation}
where the linear AR process of order $P$ is described by the filter coefficients $a_1, \dots, a_P$, while the excitation process $e_n$ is modeled by a WaveNet with a receptive field of $R$ samples.
Specifically, we assume the excitation process to be a logistic mixture
\begin{equation}
  e_n \sim \sum_{i=1}^K \pi_i \mathrm{logistic} ( \mu_i, s_i \ \vert \  e_{(n-R):(n-1)}, h_n )
  \label{eq:logistic_mixture}
\end{equation}
with non-linear dependencies to past excitation samples, as parametrized by a WaveNet. Given previous excitation samples 
$e_{(n-R):(n-1)}$
%from the time range $n-R$ to $n-1$ 
and local (acoustic) conditioning $h_n$, the WaveNet predicts the current time-step logistic mixture parameters: mixture weight $\pi_i$, component mean $\mu_i$  and component scale $s_i$. 

In this paper, the linear AR process parameters are estimated separately and kept fixed while training the excitation model. For this, we use QCP analysis, which utilizes time-weighted linear predictive analysis to attenuate the glottal contribution in the AR filter estimate \cite{Airaksinen2014}.
The linear AR process order is relatively low (we use $P$=30), whereas the receptive field of a WaveNet can grow large due to its dilated convolution structure. Furthermore, the parameters of the two processes vary at different rates: the filter parameters are updated at a 200\,Hz rate (or 5\,ms frame shift), while the excitation process parameters are predicted for every sample at a 16\,kHz rate. 

\subsection{Network architecture}

We use a WaveNet implementation based on \cite{adiga2018-wavenet-vocoder}. The model architecture has two main parts: a stack of residual blocks, which acts as a multi-scale feature extractor, and a post-processing module, which combines the information from residual blocks to predict the next signal sample. In each residual block, the key operation is a gated convolution given by
\begin{equation}
     x_\mathrm{skip} = \tanh( W_f * x_\mathrm{in} + L_f) \odot \sigma( W_g * x_\mathrm{in} + L_g),
\end{equation}
where $*$ denotes dilated causal convolution and $\odot$ is element-wise multiplication. $W_f$ and $W_g$ are convolution weight tensors for filter and gate, respectively. Additionally, $L_f$ and $L_g$ are local conditioning vectors specific to the residual block.
The skip path activations $x_\mathrm{skip}$ are connected to the post-processing module, while a residual block output $x_\mathrm{out} = W x_\mathrm{skip} + x_\mathrm{in}$ is fed forward into the next layer of the residual stack.

% \begin{equation}
%     \bm x_\mathrm{skip}  = \tanh(\bm W_f * \bm x + \bm V_f \bm h) \odot \sigma( \bm W_g * \bm x + \bm V_g \bm h)
% \end{equation}
% 

The post-processing module takes in the skip-outputs from each residual block and concatenates them along their channel dimension. This is followed by two $1 \times 1$ convolution layers with contenated rectifier activations \cite{Shang2016-concat-relu}, whose output is finally projected to the mixture density network output of size $3K$ (where $K$ is the number of mixture components).
%The final output layer size is the number of mixture components times three. 

\subsection{Local conditioning}

% TODO: move the cite to QCP to the signal model section

For local conditioning, both models use the same acoustic feature set of glottal vocoder parameters \cite{Raitio2011glotthmm}: the vocal tract filter, estimated by QCP analysis \cite{Airaksinen2014}, and the corresponding glottal source spectral envelope are parametrized by line spectrum frequencies (LSFs), using orders 30 and 10, respectively. Fundamental frequency in log-scale (LF0) and a binary voicing flag (VUV)  describe the pitch contour, whereas the average harmonic-to-noise ratio (HNR) in 5 ERB frequency-bands characterizes the signal aperiodicity. Finally, the frame energy (in dB) is used to indicate the signal level. 
%These features were chosen, as they are the default feature-set 
%These parameters are the default acoustic features in 
%
%Of these features, only the vocal tract filter is explicitly used in 

\begin{figure}[thb]
    \centering
    % trim={left, bottom, right, top}
    \includegraphics[width=\linewidth, trim={0.8cm 0.7cm 2.2cm 4.3cm}, clip]{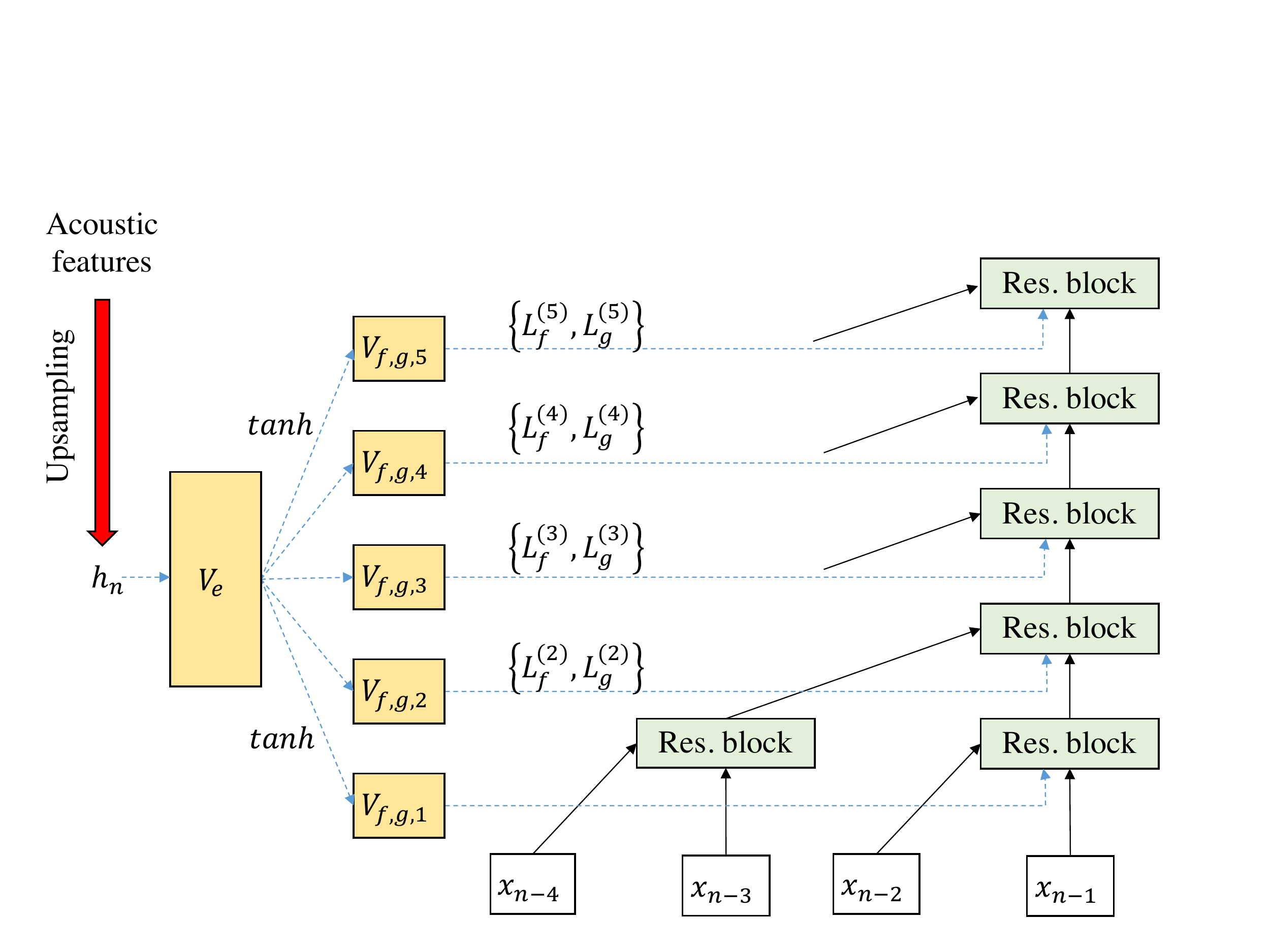}
    \caption{A five-level residual stack of a WaveNet vocoder. The residual stack shares a global embedding for the acoustic features, which is transformed to  block-specific local conditioning vectors.}
    \label{fig:wavenet_lc}
\end{figure}

In initial experiments, we found that the waveform generator reliability is improved when the model is allowed to use a small look-ahead into future conditioning. Previous work has proposed using various bi-directional recurrent structures for encoding the future of the conditioning \cite{Arik2017-deepvoice, Wang2018-comparison-of-waveform-generation}. However, training these kind of structures jointly with a WaveNet notably increases the computational cost. 
Instead, we first stack adjacent past and future frames to the current frame to provide context, after which we use linear interpolation to upsample the conditioning from 200\,Hz to 16\,Hz. Finally, we apply global projection to embed the conditioning into smaller dimensionality, before injecting the embedded conditioning into the residual modules, as shown in Fig.~\ref{fig:wavenet_lc}.
In the experiments, we use 4 frames of context to both directions, corresponding to 20\,ms look-ahead.

%Similar stacking was done in \cite{wu2015deep-multitask})

\subsection{Discretized logistic mixture density loss}

WaveNets have commonly used 8-bit quantization, which requires 256-dimensional softmax output if trained as a classifier. 
However, this often results in quantization-noise like artefacts, whereas using the full 16-bits of amplitude levels would require prohibitively large softmax layers.
%Although the observed amplitude in digital audio is quantized, the underlying physical generation mechanism is continuous. As such, it is reasonable to assume a continuous latent variable for the signal amplitude. 
To overcome this limitation, a discretized logistic mixture density loss was proposed to improve PixelCNN \cite{salimans2017-pixelcnn-plusplus}.
%(which is the image generation counterpart of WaveNet).
The approach was quickly adopted to improving WaveNet fidelity \cite{oord2017-parallel-wavenet, Shen2018-tacotron2}. 
Furthermore, mixture density networks extend more easily to multivariate modeling: for example, a WaveNet-like architecture with Gaussian mixtures has been proposed for generating vocoder parameters in singing synthesis \cite{Blaauw2017-neural-singing-synth}. 

To train a mixture density network, one has to be able to evaluate likelihoods for observations. For the logistic distribution, the cumulative distribution function (CDF) is the logistic sigmoid, and 
the probability of a quantized observation $x$ is a $\Delta$-wide slice of the CDF
\begin{equation}
    p(x) = \sum_{i=1}^K \pi_i [\sigma((x+\frac{\Delta}{2} -\mu_i) / s_i) -\sigma((x-\frac{\Delta}{2} -\mu_i) / s_i)] ,
\end{equation}
where $\Delta$ is the quantization bin width and  $\sigma$ is the logistic CDF. This formulation is then used to minimize the negative log-likelihood for the observations \cite{salimans2017-pixelcnn-plusplus}.
%
%The network is trained to minimize the negative log-likelihood of observations, given the predicted parameters.
In practice, the network outputs are treated as mixture weight logits, 
%(where a softmax normalization is applied),
component means and log-scale parameters. Notably, the log-scales should be floored to avoid variance collapse, but the floor level simultaneously acts as a noise floor in generation. If the floor is set too high, this property may lead to exaggerated background noise or roughness in the synthetic voiced speech.

\section{Experiments}
\label{sec:experiments}

\subsection{Speech material}

%the name of my font is \familydefault

We use a multi-speaker database originally released for speech enhancement research \cite{Valentini-Botinhao2017-noisy-speech-database}, and only take the clean speech subset for these experiments. The voice talents in the dataset are non-professional native British English speakers. % Is this correct?
The full training dataset consists of 56 speakers, but to scale the task for our available computational resources, we use a 28-speaker subset provided in the data. We treat these data as our seen speakers dataset, which contains 11571 utterances in total, amounting to 9.4 hours of speech, i.e., about 20 minutes per seen speaker. The ten first utterances from each seen speaker were reserved for testing, and 500 of the remaining utterances were randomly chosen for validation. 
Additionally, two speakers (one female, one male) from the database testset were held out as unseen.

% Training set total duration 33805.2 seconds

\subsection{Training the models}

For both WaveNet and GlotNet, we used 64 channels within the residual blocks (residual and skip channels) and 128 channels in the post-processing module. The convolution filter width is two everywhere in the residual stack, in which the dilation pattern $1,2,4,\dots,512$ is  repeated three times, resulting in a total of 30 residual blocks and a receptive field length of 3071 samples.
The training criterion for the models was to minimize 
the discretized logistic mixture negative log-likelihood for their respective observed signals, where we used 5 mixture components. The models were trained for 70 epochs (with a 10 epoch early stopping criterion) using the Adam optimizer \cite{Kingma2014-adam} and exponential moving average weight smoothing \cite{Polyak1992-EMA}.

%
%Mixture density network seems more difficult to train than a softmax
The prediction of signal sample probability distributions allows manual adjustment of the sampling strategy at test time, for example, mode sampling in voiced regions has been reported to improve perceived synthetic speech quality \cite{Wang2018-comparison-of-waveform-generation}.
Nevertheless, we chose to sample directly from the predicted distributions as we feel this accurately reflects the learned model quality.

\subsection{Listening tests}

For subjective evaluation of the system performances, we conducted listening tests on speaker similarity and speech quality.%
\footnote{
Samples available at \url{https://users.aalto.fi/~ljuvela/is18_glotnet/} 
}
The tests were run on the CrowdFlower crowd-sourcing platform \cite{crowdflower}, where the tests were made available in English-speaking countries and  the top four countries in the  EFI English proficiency rating \cite{efi-english-proficiency-index}.
Each test case was evaluated by 50 listeners, while the listeners were screened using natural reference null pairs and artificially corrupted anchor samples. 

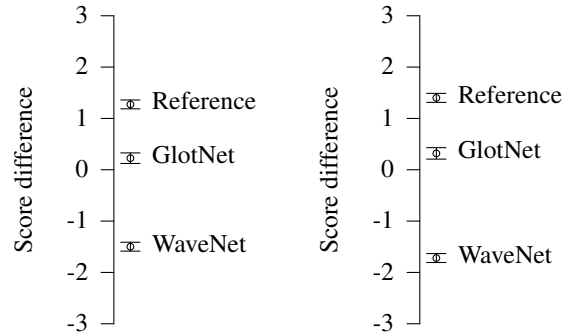
\begin{figure}[htb]
\centering
{
\large
\resizebox{0.49\linewidth}{!}{\begin{tikzpicture}[x=1pt,y=1pt]
\definecolor[named]{fillColor}{rgb}{1.00,1.00,1.00}
\path[use as bounding box,fill=fillColor,fill opacity=0.00] (0,0) rectangle (108.41,144.54);
\begin{scope}
\path[clip] (  0.00,  0.00) rectangle (108.41,144.54);
\definecolor[named]{drawColor}{rgb}{0.00,0.00,0.00}

\path[draw=drawColor,line width= 0.4pt,line join=round,line cap=round] ( 54.19, 96.90) circle (  1.26);

\path[draw=drawColor,line width= 0.4pt,line join=round,line cap=round] ( 54.19, 76.63) circle (  1.26);

\path[draw=drawColor,line width= 0.4pt,line join=round,line cap=round] ( 54.19, 43.28) circle (  1.26);

\path[draw=drawColor,line width= 0.4pt,line join=round,line cap=round] ( 48.00, 14.24) -- ( 48.00,130.30);

\path[draw=drawColor,line width= 0.4pt,line join=round,line cap=round] ( 48.00, 14.24) -- ( 43.20, 14.24);

\path[draw=drawColor,line width= 0.4pt,line join=round,line cap=round] ( 48.00, 33.58) -- ( 43.20, 33.58);

\path[draw=drawColor,line width= 0.4pt,line join=round,line cap=round] ( 48.00, 52.93) -- ( 43.20, 52.93);

\path[draw=drawColor,line width= 0.4pt,line join=round,line cap=round] ( 48.00, 72.27) -- ( 43.20, 72.27);

\path[draw=drawColor,line width= 0.4pt,line join=round,line cap=round] ( 48.00, 91.61) -- ( 43.20, 91.61);

\path[draw=drawColor,line width= 0.4pt,line join=round,line cap=round] ( 48.00,110.96) -- ( 43.20,110.96);

\path[draw=drawColor,line width= 0.4pt,line join=round,line cap=round] ( 48.00,130.30) -- ( 43.20,130.30);

\node[text=drawColor,anchor=base east,inner sep=0pt, outer sep=0pt, scale=  0.80] at ( 38.40, 11.49) {-3};

\node[text=drawColor,anchor=base east,inner sep=0pt, outer sep=0pt, scale=  0.80] at ( 38.40, 30.83) {-2};

\node[text=drawColor,anchor=base east,inner sep=0pt, outer sep=0pt, scale=  0.80] at ( 38.40, 50.17) {-1};

\node[text=drawColor,anchor=base east,inner sep=0pt, outer sep=0pt, scale=  0.80] at ( 38.40, 69.52) {0};

\node[text=drawColor,anchor=base east,inner sep=0pt, outer sep=0pt, scale=  0.80] at ( 38.40, 88.86) {1};

\node[text=drawColor,anchor=base east,inner sep=0pt, outer sep=0pt, scale=  0.80] at ( 38.40,108.20) {2};

\node[text=drawColor,anchor=base east,inner sep=0pt, outer sep=0pt, scale=  0.80] at ( 38.40,127.54) {3};

\node[text=drawColor,rotate= 90.00,anchor=base,inner sep=0pt, outer sep=0pt, scale=  0.80] at ( 17.28, 72.27) {Score difference};

\path[draw=drawColor,line width= 0.4pt,line join=round,line cap=round] ( 54.19, 95.23) -- ( 54.19, 98.57);

\path[draw=drawColor,line width= 0.4pt,line join=round,line cap=round] ( 50.57, 95.23) --
	( 54.19, 95.23) --
	( 57.80, 95.23);

\path[draw=drawColor,line width= 0.4pt,line join=round,line cap=round] ( 57.80, 98.57) --
	( 54.19, 98.57) --
	( 50.57, 98.57);

\path[draw=drawColor,line width= 0.4pt,line join=round,line cap=round] ( 54.19, 74.64) -- ( 54.19, 78.62);

\path[draw=drawColor,line width= 0.4pt,line join=round,line cap=round] ( 50.57, 74.64) --
	( 54.19, 74.64) --
	( 57.80, 74.64);

\path[draw=drawColor,line width= 0.4pt,line join=round,line cap=round] ( 57.80, 78.62) --
	( 54.19, 78.62) --
	( 50.57, 78.62);

\path[draw=drawColor,line width= 0.4pt,line join=round,line cap=round] ( 54.19, 41.61) -- ( 54.19, 44.96);

\path[draw=drawColor,line width= 0.4pt,line join=round,line cap=round] ( 50.57, 41.61) --
	( 54.19, 41.61) --
	( 57.80, 41.61);

\path[draw=drawColor,line width= 0.4pt,line join=round,line cap=round] ( 57.80, 44.96) --
	( 54.19, 44.96) --
	( 50.57, 44.96);

\node[text=drawColor,anchor=base west,inner sep=0pt, outer sep=0pt, scale=  0.80] at ( 62.33, 95.06) {Reference};

\node[text=drawColor,anchor=base west,inner sep=0pt, outer sep=0pt, scale=  0.80] at ( 62.33, 74.79) {GlotNet};

\node[text=drawColor,anchor=base west,inner sep=0pt, outer sep=0pt, scale=  0.80] at ( 62.33, 41.45) {WaveNet};
\end{scope}
\end{tikzpicture}}
\resizebox{0.49\linewidth}{!}{\begin{tikzpicture}[x=1pt,y=1pt]
\definecolor[named]{fillColor}{rgb}{1.00,1.00,1.00}
\path[use as bounding box,fill=fillColor,fill opacity=0.00] (0,0) rectangle (108.41,144.54);
\begin{scope}
\path[clip] (  0.00,  0.00) rectangle (108.41,144.54);
\definecolor[named]{drawColor}{rgb}{0.00,0.00,0.00}

\path[draw=drawColor,line width= 0.4pt,line join=round,line cap=round] ( 54.19, 99.36) circle (  1.26);

\path[draw=drawColor,line width= 0.4pt,line join=round,line cap=round] ( 54.19, 78.43) circle (  1.26);

\path[draw=drawColor,line width= 0.4pt,line join=round,line cap=round] ( 54.19, 39.01) circle (  1.26);

\path[draw=drawColor,line width= 0.4pt,line join=round,line cap=round] ( 48.00, 14.24) -- ( 48.00,130.30);

\path[draw=drawColor,line width= 0.4pt,line join=round,line cap=round] ( 48.00, 14.24) -- ( 43.20, 14.24);

\path[draw=drawColor,line width= 0.4pt,line join=round,line cap=round] ( 48.00, 33.58) -- ( 43.20, 33.58);

\path[draw=drawColor,line width= 0.4pt,line join=round,line cap=round] ( 48.00, 52.93) -- ( 43.20, 52.93);

\path[draw=drawColor,line width= 0.4pt,line join=round,line cap=round] ( 48.00, 72.27) -- ( 43.20, 72.27);

\path[draw=drawColor,line width= 0.4pt,line join=round,line cap=round] ( 48.00, 91.61) -- ( 43.20, 91.61);

\path[draw=drawColor,line width= 0.4pt,line join=round,line cap=round] ( 48.00,110.96) -- ( 43.20,110.96);

\path[draw=drawColor,line width= 0.4pt,line join=round,line cap=round] ( 48.00,130.30) -- ( 43.20,130.30);

\node[text=drawColor,anchor=base east,inner sep=0pt, outer sep=0pt, scale=  0.80] at ( 38.40, 11.49) {-3};

\node[text=drawColor,anchor=base east,inner sep=0pt, outer sep=0pt, scale=  0.80] at ( 38.40, 30.83) {-2};

\node[text=drawColor,anchor=base east,inner sep=0pt, outer sep=0pt, scale=  0.80] at ( 38.40, 50.17) {-1};

\node[text=drawColor,anchor=base east,inner sep=0pt, outer sep=0pt, scale=  0.80] at ( 38.40, 69.52) {0};

\node[text=drawColor,anchor=base east,inner sep=0pt, outer sep=0pt, scale=  0.80] at ( 38.40, 88.86) {1};

\node[text=drawColor,anchor=base east,inner sep=0pt, outer sep=0pt, scale=  0.80] at ( 38.40,108.20) {2};

\node[text=drawColor,anchor=base east,inner sep=0pt, outer sep=0pt, scale=  0.80] at ( 38.40,127.54) {3};

\node[text=drawColor,rotate= 90.00,anchor=base,inner sep=0pt, outer sep=0pt, scale=  0.80] at ( 17.28, 72.27) {Score difference};

\path[draw=drawColor,line width= 0.4pt,line join=round,line cap=round] ( 54.19, 97.66) -- ( 54.19,101.06);

\path[draw=drawColor,line width= 0.4pt,line join=round,line cap=round] ( 50.57, 97.66) --
	( 54.19, 97.66) --
	( 57.80, 97.66);

\path[draw=drawColor,line width= 0.4pt,line join=round,line cap=round] ( 57.80,101.06) --
	( 54.19,101.06) --
	( 50.57,101.06);

\path[draw=drawColor,line width= 0.4pt,line join=round,line cap=round] ( 54.19, 76.28) -- ( 54.19, 80.59);

\path[draw=drawColor,line width= 0.4pt,line join=round,line cap=round] ( 50.57, 76.28) --
	( 54.19, 76.28) --
	( 57.80, 76.28);

\path[draw=drawColor,line width= 0.4pt,line join=round,line cap=round] ( 57.80, 80.59) --
	( 54.19, 80.59) --
	( 50.57, 80.59);

\path[draw=drawColor,line width= 0.4pt,line join=round,line cap=round] ( 54.19, 37.33) -- ( 54.19, 40.69);

\path[draw=drawColor,line width= 0.4pt,line join=round,line cap=round] ( 50.57, 37.33) --
	( 54.19, 37.33) --
	( 57.80, 37.33);

\path[draw=drawColor,line width= 0.4pt,line join=round,line cap=round] ( 57.80, 40.69) --
	( 54.19, 40.69) --
	( 50.57, 40.69);

\node[text=drawColor,anchor=base west,inner sep=0pt, outer sep=0pt, scale=  0.80] at ( 62.33, 97.53) {Reference};

\node[text=drawColor,anchor=base west,inner sep=0pt, outer sep=0pt, scale=  0.80] at ( 62.33, 76.60) {GlotNet};

\node[text=drawColor,anchor=base west,inner sep=0pt, outer sep=0pt, scale=  0.80] at ( 62.33, 37.18) {WaveNet};
\end{scope}
\end{tikzpicture}}
}
\caption{Combined score differences obtained from the quality comparison CCR test for seen speakers (left) and unseen speakers (right). Error bars are t-statistic based 95\% confidence intervals for the mean.}
\label{fig:CCR_rankings}
\end{figure}

To evaluate the subjective quality of the synthetic speech, we conducted pairwise category comparison rating (CCR) tests \cite{Itu1996}, where the listeners were presented with a pair of samples and asked to rate the comparative quality on a 7-level scale, ranging from -3 (much worse) to 3 (much better).
Combined scores are shown in Fig.~\ref{fig:CCR_rankings}. The scores were calculated by reordering the ratings for each system and pooling together all ratings the system received. Natural speech target utterance was included in the tests as a reference system. The plots show mean ratings with 95\% confidence, corrected for multiple comparisons.

\begin{figure}[htb]
\centering
{
\large
\includegraphics[height=0.75\linewidth]{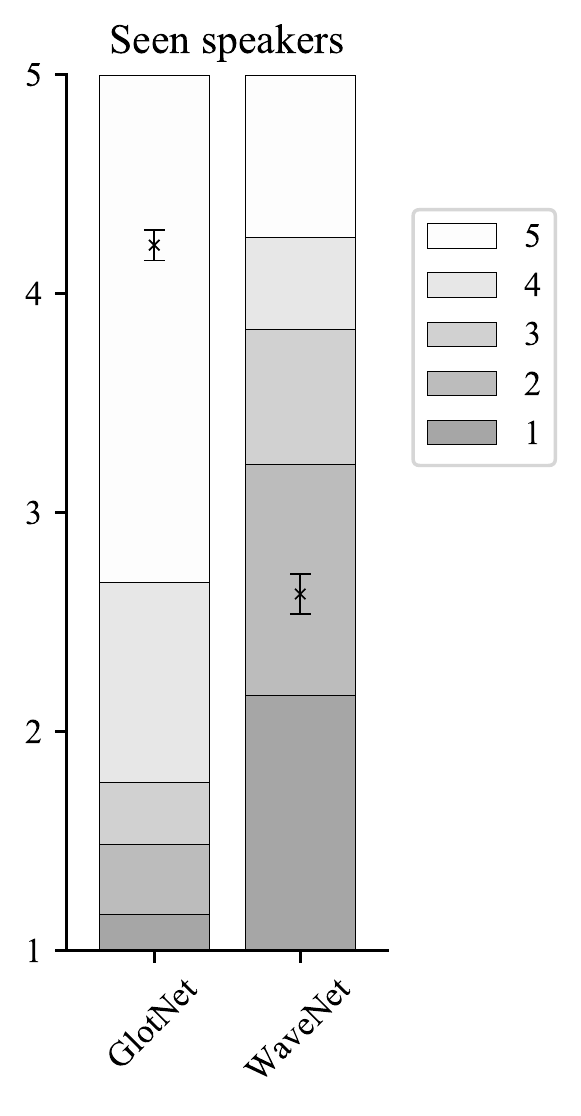}
\includegraphics[height=0.75\linewidth]{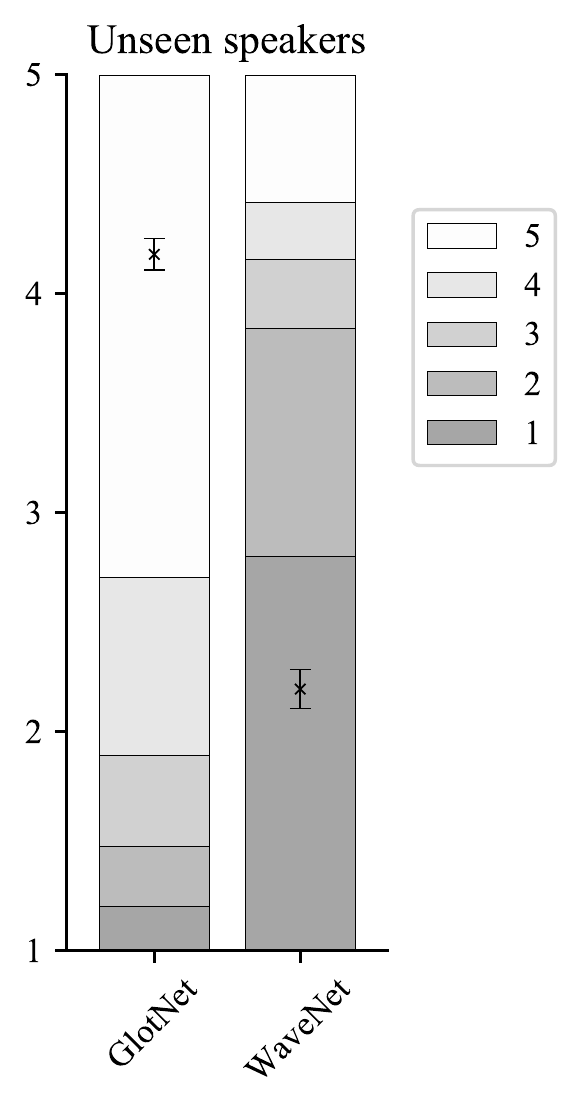}
%\resizebox{0.49\linewidth}{!}{\input{figures/score_distributions_unseen.tex}}
}
\caption{Voice similarity ratings in a DMOS test for seen speakers (left) and unseen speakers (right). Mean scores are shown with 95\% confidence intervals, while  relative score distribution histograms are shown in the background.}
\label{fig:DMOS_rankings}
\end{figure}

Synthetic speech voice similarity to a natural reference was measured in a DMOS-like  test \cite{Itu1996}. The listeners were presented with a test sample and asked to rate the voice similarity to the target natural speech utterance on a 5-level absolute category rating scale, ranging from 1 (bad) to 5 (excellent). Results are shown in Fig.~\ref{fig:DMOS_rankings}. The plot shows mean ratings with 95\% confidence intervals, as well as stacked score distribution histograms in the background.

In both test types, GlotNet performs favourably to WaveNet. Furthermore, GlotNet ratings remain largely unaffected by testing on unseen speakers, whereas WaveNet scores slightly decrease. It should be noted that both tests involve paired comparisons to a natural speech reference, which makes the tests quite sensitive to small degradations.

\subsection{Objective measures}

\begin{figure}
    \centering
    {
    \normalsize
    \resizebox{\linewidth}{!}{\input{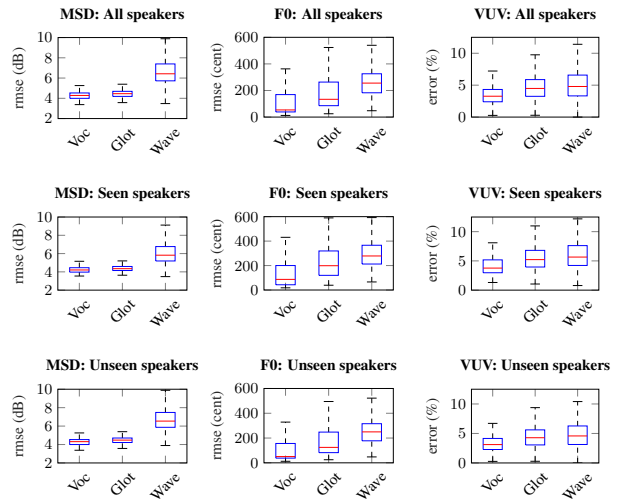}}
    }
    \caption{Objective measures for mel spectral distortion (MSD), log-F0 RMSE (in cents) and voicing decision error (\%). 'Voc' denotes a deterministic glottal vocoder, while 'Glot' and 'Wave' are GlotNet and WaveNet vocoders, respectively.}
    \label{fig:objective_measures}
\end{figure}

To quantify how reliably the different waveform generation methods follow their acoustic conditioning, we computed various objective metrics. 
%These measures are not equivalent with the perceptual quality, but estimate how closely the synthesis method follows the acoustic conditioning.
Fig.~\ref{fig:objective_measures} shows objective measures for different systems, computed with respect to the original signal. The box-and-whiskers plots show the medians, along with the 25\% and 75\% quantiles.
% can also be inner an outer quartiles
A deterministic glottal vocoder which uses the same acoustic feature set is included as a reference method.
Mel spectral distortion (MCD, in dB) was calculated by applying a 24-band mel filterbank matrix to FFT magnitude spectrum, and taking the root-mean-squared error of the log-differences over frames and mel-bands. 
F0 was estimated from the synthetic signals using the RAPT algorithm \cite{talkin1995-rapt}, and log-domain 
F0 difference (in cents: 100 cents is one semitone, 12 semitones is one octave) is reported over frames where the voicing estimates agree. Finally, we report the voicing error percentages between the local conditioning and the one estimated from the synthetic signals.

% \begin{equation}
%     \mathrm{MSD}(x_1, x_2) = 10 \sum_{n=1}^N \frac{1}{N} ( \log_{10}( \bm M X_{1,n} ) - \log_{10}( \bm M X_{2,n} ) )^2    
% \end{equation}

%
% \begin{table}[htb]
%   \caption{Objective measures}
%   \label{tab:word_styles}
%   \centering
%   \begin{tabular}{lllll}
%     \toprule
%      &
%     \multicolumn{2}{l}{F0 dist (cent)} &
%     \multicolumn{2}{l}{MFCC dist (dB)} \\
%     \textbf{Method} & seen & unseen & seen & unseen \\
%     \midrule
%     GlotNet   & Foo & Bar & Foo & Bar \\
%     WaveNet   & Foo & Bar & Foo & Bar \\
%     \bottomrule
%   \end{tabular}
% \end{table}

\section{Discussion}
\label{sec:discussion}

In the present experiments, the direct waveform WaveNet vocoder performance appears lacking both in terms of subjective quality and objective reliability. This can be largely attributed to the  multi-speaker task combined with the relatively small dataset and computation budget. Furthermore, we feel that the logistic mixture density network training is more demanding than the softmax-based approach. Previously, high-quality logistic mixture WaveNets have been trained using more data and speaker-specific models \cite{oord2017-parallel-wavenet, Shen2018-tacotron2}, whereas previous speaker-independent models have used the softmax training approach \cite{Hayashi2017-multispeaker-wavenet-vocoder, Kleijn2018-wavenet-low-rate-coding}. We also note that our models use relatively few parameters compared to previous research. As such,
the WaveNet vocoder performance would likely improve by using more training data and larger models. 

Nevertheless, adding the low-order linear AR component to the signal model in GlotNet considerably improves the model performance with the same data and equivalent model architecture and training procedure. This is well motivated by the prevalent use of linear predictive models in speech applications. Furthermore, GlotNet-like excitation models should be well applicable to existing parametric TTS systems, as their acoustic features often include spectral envelope information interpretable as a filter. Among these spectral features, glottal inverse filtering based models are physiologically motivated and aim to consistently separate the excitation signal from the linear AR envelope filter. 

% In related work, the softmax approach has been used with good results to create speaker-specific WaveNet vocoder models with equivalent acoustic features \cite{Juvela2018-glotnet-tts}.

\section{Conclusions}
\label{sec:conclusions}

This paper proposed a speaker-independent neural waveform generator  which combines a linear autoregressive (vocal tract filter) process  with a non-linear (glottal source) excitation process parametrized by a WaveNet. Listening tests and objective measures show that the proposed method outperforms directly modeling speech with a WaveNet vocoder, when both models use identical architectures and training data.
While the current work focuses on copy-synthesis experiments, future work includes integrating the waveform generator models into parametric text-to-speech systems.

\section{Acknowledgements}
This work was supported by the Academy of Finland (proj.~no.~284671 and 312490) and MEXT KAKENHI Grant Numbers (15H01686, 16H06302, 17H04687).
We acknowledge the computational resources provided by the Aalto Science-IT project.

\newpage

\bibliographystyle{IEEEtran}

\bibliography{refs}

\begin{thebibliography}{10}
\providecommand{\url}[1]{#1}
\csname url@samestyle\endcsname
\providecommand{\newblock}{\relax}
\providecommand{\bibinfo}[2]{#2}
\providecommand{\BIBentrySTDinterwordspacing}{\spaceskip=0pt\relax}
\providecommand{\BIBentryALTinterwordstretchfactor}{4}
\providecommand{\BIBentryALTinterwordspacing}{\spaceskip=\fontdimen2\font plus
\BIBentryALTinterwordstretchfactor\fontdimen3\font minus
  \fontdimen4\font\relax}
\providecommand{\BIBforeignlanguage}[2]{{%
\expandafter\ifx\csname l@#1\endcsname\relax
\typeout{** WARNING: IEEEtran.bst: No hyphenation pattern has been}%
\typeout{** loaded for the language `#1'. Using the pattern for}%
\typeout{** the default language instead.}%
\else
\language=\csname l@#1\endcsname
\fi
#2}}
\providecommand{\BIBdecl}{\relax}
\BIBdecl

\bibitem{oord2016-wavenet}
\BIBentryALTinterwordspacing
A.~van~den Oord, S.~Dieleman, H.~Zen, K.~Simonyan, O.~Vinyals, A.~Graves,
  N.~Kalchbrenner, A.~Senior, and K.~Kavukcuoglu, ``{WaveNet}: A generative
  model for raw audio,'' \emph{arXiv pre-print}, 2016. [Online]. Available:
  \url{https://arxiv.org/pdf/1609.03499}
\BIBentrySTDinterwordspacing

\bibitem{Arik2017-deepvoice}
S.~O. Arik, M.~Chrzanowski, A.~Coates, G.~Diamos, A.~Gibiansky, Y.~Kang, X.~Li,
  J.~Miller, A.~Ng, J.~Raiman, S.~Sengupta, and M.~Shoeybi, ``{Deep Voice}:
  Real-time neural text-to-speech,'' in \emph{Proc.~ICML}, 2017.

\bibitem{Shen2018-tacotron2}
J.~Shen, R.~Pang, R.~Weiss, M.~Schuster, N.~Jaitly, Z.~Yang, Z.~Chen, Y.~Zhang,
  Y.~Wang, R.~Skerry-Ryan, R.~Saurous, Y.~Agiomyrgiannakis, and Y.~Wu,
  ``Natural {TTS} synthesis by conditioning {WaveNet} on mel spectrogram
  predictions,'' in \emph{Proc.~ICASSP}, 2018.

\bibitem{Tamamori2017-wavenet-vocoder}
A.~Tamamori, T.~Hayashi, K.~Kobayashi, K.~Takeda, and T.~Toda,
  ``Speaker-dependent {WaveNet} vocoder,'' in \emph{Proc.~Interspeech}, 2017,
  pp. 1118--1122.

\bibitem{Wang2018-comparison-of-waveform-generation}
X.~Wang, J.~Lorenzo-Trueba, S.~Takaki, L.~Juvela, and J.~Yamagishi, ``A
  comparison of recent waveform generation and acoustic modeling methods for
  neural-network-based speech synthesis,'' in \emph{Proc.~ICASSP}, 2018.

\bibitem{Kleijn2018-wavenet-low-rate-coding}
W.~B. Kleijn, F.~S.~C. Lim, A.~Luebs, J.~Skoglund, F.~Stimberg, Q.~Wang, and
  T.~C. Walters, ``Wavenet based low rate speech coding,'' in
  \emph{Proc.~ICASSP}, 2018.

\bibitem{Hayashi2017-multispeaker-wavenet-vocoder}
T.~Hayashi, A.~Tamamori, K.~Kobayashi, K.~Takeda, and T.~Toda, ``An
  investigation of multi-speaker training for {WaveNet} vocoder,'' in
  \emph{Proc.~ASRU}, Dec 2017, pp. 712--718.

\bibitem{juvela2016a-high-pitched-excitation}
L.~Juvela, B.~Bollepalli, M.~Airaksinen, and P.~Alku, ``High-pitched excitation
  generation for glottal vocoding in statistical parametric speech synthesis
  using a deep neural network,'' in \emph{Proc.~ICASSP}, March 2016, pp.
  5120--5124.

\bibitem{airaksinen2016glottdnn}
M.~Airaksinen, B.~Bollepalli, L.~Juvela, Z.~Wu, S.~King, and P.~Alku,
  ``{GlottDNN}---a full-band glottal vocoder for statistical parametric speech
  synthesis,'' in \emph{Proc.~Interspeech}, 2016.

\bibitem{Airaksinen2014}
M.~Airaksinen, T.~Raitio, B.~Story, and P.~Alku, ``Quasi closed phase glottal
  inverse filtering analysis with weighted linear prediction,'' \emph{IEEE/ACM
  Transactions on Audio, Speech, and Language Processing}, vol.~22, no.~3, pp.
  596--607, March 2014.

\bibitem{Alku2011}
P.~Alku, ``Glottal inverse filtering analysis of human voice production -- a
  review of estimation and parameterization methods of the glottal excitation
  and their applications. (invited article),'' \emph{Sadhana -- Academy
  Proceedings in Engineering Sciences}, vol.~36, no.~5, pp. 623--650, 2011.

\bibitem{Bollepalli2017-gan-glottal-excitation}
B.~Bollepalli, L.~Juvela, and P.~Alku, ``Generative adversarial network-based
  glottal waveform model for statistical parametric speech synthesis,'' in
  \emph{Proc.~Interspeech}, 2017, pp. 3394--3398.

\bibitem{juvela2018-synthesis-from-mfcc}
L.~Juvela, B.~Bollepalli, X.~Wang, H.~Kameoka, M.~Airaksinen, J.~Yamagishi, and
  P.~Alku, ``Speech waveform synthesis from {MFCC} sequences with generative
  adversarial networks,'' in \emph{Proc.~ICASSP}, 2018.

\bibitem{adiga2018-wavenet-vocoder}
N.~Adiga, V.~Tsiaras, and Y.~Stylianou, ``On the use of {WaveNet} as a
  statistical vocoder,'' in \emph{Proc.~of ICASSP}, 2018.

\bibitem{Shang2016-concat-relu}
\BIBentryALTinterwordspacing
W.~Shang, K.~Sohn, D.~Almeida, and H.~Lee, ``Understanding and improving
  convolutional neural networks via concatenated rectified linear units,''
  \emph{arXiv pre-print}, 2016. [Online]. Available:
  \url{http://arxiv.org/abs/1603.05201}
\BIBentrySTDinterwordspacing

\bibitem{Raitio2011glotthmm}
T.~Raitio, A.~Suni, J.~Yamagishi, H.~Pulakka, J.~Nurminen, M.~Vainio, and
  P.~Alku, ``{HMM}-based speech synthesis utilizing glottal inverse
  filtering,'' \emph{IEEE Transactions on Audio, Speech, and Language
  Processing}, vol.~19, no.~1, pp. 153--165, January 2011.

\bibitem{salimans2017-pixelcnn-plusplus}
\BIBentryALTinterwordspacing
T.~Salimans, A.~Karpathy, X.~Chen, and D.~P. Kingma, ``{PixelCNN++}: Improving
  the {PixelCNN} with discretized logistic mixture likelihood and other
  modifications,'' \emph{arXiv pre-print}, 2017. [Online]. Available:
  \url{http://arxiv.org/abs/1701.05517}
\BIBentrySTDinterwordspacing

\bibitem{oord2017-parallel-wavenet}
\BIBentryALTinterwordspacing
A.~van~den Oord, Y.~Li, I.~Babuschkin, K.~Simonyan, O.~Vinyals, K.~Kavukcuoglu,
  G.~van~den Driessche, E.~Lockhart, L.~C. Cobo, F.~Stimberg, N.~Casagrande,
  D.~Grewe, S.~Noury, S.~Dieleman, E.~Elsen, N.~Kalchbrenner, H.~Zen,
  A.~Graves, H.~King, T.~Walters, D.~Belov, and D.~Hassabis, ``Parallel
  {WaveNet}: Fast high-fidelity speech synthesis,'' \emph{arXiv pre-print},
  2017. [Online]. Available: \url{http://arxiv.org/abs/1711.10433}
\BIBentrySTDinterwordspacing

\bibitem{Blaauw2017-neural-singing-synth}
M.~Blaauw and J.~Bonada, ``A neural parametric singing synthesizer,'' in
  \emph{Proc.~Interspeech}, 2017, pp. 4001--4005.

\bibitem{Valentini-Botinhao2017-noisy-speech-database}
\BIBentryALTinterwordspacing
C.~Valentini-Botinhao, ``Noisy speech database for training speech enhancement
  algorithms and {TTS} models,'' 2017. [Online]. Available:
  \url{http://dx.doi.org/10.7488/ds/2117}
\BIBentrySTDinterwordspacing

\bibitem{Kingma2014-adam}
\BIBentryALTinterwordspacing
D.~P. Kingma and J.~Ba, ``Adam: {A} method for stochastic optimization,'' in
  \emph{Proc.~ICLR}, 2015. [Online]. Available:
  \url{http://arxiv.org/abs/1412.6980}
\BIBentrySTDinterwordspacing

\bibitem{Polyak1992-EMA}
B.~T. Polyak and A.~B. Juditsky, ``Acceleration of stochastic approximation by
  averaging,'' \emph{SIAM J. Control Optimization}, vol.~30, no.~4, pp.
  838--855, 1992.

\bibitem{crowdflower}
{CrowdFlower Inc.}, ``Crowd-sourcing platform,'' https://www.crowdflower.com/,
  accessed: 2018-03-22.

\bibitem{efi-english-proficiency-index}
``{EF} {E}nglish proficiency index,'' http://www.ef.com/epi/, accessed:
  2018-03-22.

\bibitem{Itu1996}
``{Methods for Subjective Determination of Transmission Quality},'' ITU-T SG12,
  Geneva, Switzerland, Recommendation P.800, Aug. 1996.

\bibitem{talkin1995-rapt}
D.~Talkin, ``A robust algorithm for pitch tracking ({RAPT}),'' \emph{Speech
  coding and synthesis}, vol. 495, p. 518, 1995.

\end{thebibliography}

\end{document}